\documentclass[manuscript]
{acmart}

\setcopyright{none}

\usepackage{math-setup}

\begin{document}

\title{\toolname: Synthesis-Powered Multi-Modal Image Search}

\author{Celeste Barnaby}
\orcid{0000-0001-7688-6133}
\affiliation{
  \institution{University of Texas at Austin}
  \country{USA} 
}
\email{celestebarnaby@utexas.edu} 

\author{Qiaochu Chen}
\orcid{0000-0003-4680-5157}             
\affiliation{           
  \institution{University of Texas at Austin}           
  \country{USA}                   
}
\email{qchen@cs.utexas.edu}

\author{Chenglong Wang}
\orcid{0000-0002-5933-6620}             
\affiliation{           
  \institution{Microsoft Research}           
  \country{USA}                   
}
\email{chenwang@microsoft.com}

\author{Işıl Dillig}
\orcid{0000-0001-8006-1230}
\affiliation{
    \institution{University of Texas at Austin}
    \country{USA}
}
\email{isil@cs.utexas.edu}


\begin{abstract}
Due to the availability of increasingly large amounts of visual data, there is a growing need for tools that can help users  find relevant images. While existing tools
can perform image retrieval based on similarity or metadata, they fall short in scenarios that necessitate semantic reasoning about the  content of the image. This paper explores a new \emph{multi-modal image search approach} that allows users to conveniently specify and perform semantic image search tasks.
With our tool, \toolname, the user interactively provides natural language descriptions, positive and negative examples, and object tags to specify their search tasks. Under the hood, \toolname\ is powered by a program synthesis engine that generates visual queries in a domain-specific language and executes the synthesized program to retrieve the desired images. In a study with 25 participants, we observed that \toolname\ allows users to perform image retrieval tasks more accurately and with less manual effort.


\end{abstract}

\maketitle

\section{Introduction}\label{sec:intro}



With the advancement of camera technologies and the prevalence of social media, photography is more accessible than ever. Nowadays, people increasingly have access to large volumes of photographs, taken by themselves or shared by others, that capture unique moments of their lives. As this volume grows, the task of  retrieving relevant images from one's personal library becomes more important yet also more challenging. Modern photo management tools like Google Photos allow the user to search for relevant images based on metadata constraints (e.g., presence of a specific person; the date the photo was taken) or visual similarity to another image or natural language query.

While  existing interfaces work reasonably well for simple search tasks, they fall short in \emph{structured image retrieval tasks}: that is, tasks that require semantic reasoning about the structure of objects in an image. Such structured image retrieval tasks are important both for professional photographers as well as regular users who increasingly have access to large amounts of visual data on their smartphones and the cloud.  For example, event photographers often have a shot list describing certain images that they must deliver to a client, such as those where the bride and groom are walking down the aisle  or  images containing only the bride and her mother \cite{weddingshotlist, weddingblog}. However, such structured image search tasks also come up in everyday life for regular users. For example, someone who is writing a travel blog might want to retrieve those images in which they are standing in front of the Eiffel  Tower, or someone mourning the loss of a pet might want to find all images in which their cat is sitting on their lap. 

As illustrated by these examples, such structured image search tasks require reasoning about contents of the image as well as relationships between them. However, such tasks are not easy to specify using existing image search interfaces. For example, while they provide support for finding images that contain a specific person, they do not facilitate searching for images where that person is performing a certain action or has a certain property. In fact, a key characteristic of structured image search tasks is that they require the contents of the image to satisfy certain \emph{logical constraints} and combinations thereof.

In this paper, we propose a new user  interaction model that facilitates structured image search. In general, these structured image retrieval tasks pose two challenges: First, how can a user effectively communicate their intent to the image search tool? Second, how can the search tool plan and execute the search logic underlying the user's intent? 

\begin{itemize}[leftmargin=*]
 \item {\bf User specification challenge: } For some image search tasks, it is difficult for users to convey their intent with a single modality. In particular, an example image alone is often too ambiguous to convey complex search logic. On the other hand, natural language (NL) alone also has shortcomings. For instance, even ostensibly simple relational attributes like ``next to'' or ``on top of'' can have multiple possible interpretations that are difficult to disambiguate without a visual example. 
 
 \item {\bf System development challenge:} Existing text or object-based image search tools are powered by vision-language models ~\cite{clip, flamingo}. Despite their object understanding capabilities, they have limited capability in reasoning about complex search logic involving multiple constraints and semantic relationships between different objects. Hence, even if the user is able to perfectly convey their intent, there are no existing techniques that can be used to execute complex image retrieval queries.
 
\end{itemize}

We propose to address the above challenges of structured image retrieval through a novel program synthesis-powered \emph{multi-modal image search tool}, \toolname{}. With \toolname{}, users can communicate their intent using a combination of natural language, positive and negative example images, and interactive object tagging. Through \toolname's multi-modal specification interface, the user can start with an efficient NL description of the task, then iteratively refine the search results by responding to queries posed through this interface. Under the hood, \toolname's backend synthesizer generates a programmatic query expressed in a domain-specific language (DSL) for image retrieval. If the user's NL description is ambiguous, the generated program will be incomplete, allowing \toolname{} to ask clarifying questions to the user in a goal-directed way. Once all ambiguities are resolved through user interaction and \toolname{} generates  a complete program, the resulting query is executed on all uploaded images and search results are displayed to the user. At that point, the user can inspect the search results and further refine them if  needed.

To assess the efficacy of \toolname{} compared to  alternatives, we have conducted a user study involving 25 participants. We find that, compared with a baseline image search tool (leveraging a state-of-the-art vision model), users see a 34\% increase in the F1 score of their search results when using \toolname{}. Further, in post-study interviews, users report that they are better able to convey their intent via \toolname{}'s multi-modal specification interface and have more trust that \toolname{}'s results are correct.


To summarize, this paper makes the following contributions: 
\begin{enumerate}
\item We present a new  multi-modal image search interface targeted towards \emph{structured} image retrieval tasks that allows users to effectively communicate their intent in an interactive fashion.
\item We describe a neuro-symbolic image query language that allows expressing the types of logical queries that underlie structured image retrieval tasks.
\item We present a program synthesis technique that leverages all the different modalities of input that  users can provide through our proposed interface.
\end{enumerate}

\section{Related Work}\label{sec:related}

\subsection{Image Retrieval}

\toolname{} performs content-based image retrieval (CBIR), a technology pivotal in organizing digital image archives by  visual content~\cite{cbir_00_survey}. \citet{cbir_08_survey} characterize CBIR tools from two perspectives: the user's and the system's. The user perspective  depends on input query modalities, while the system perspective hinges on query processing methods and presentation of search results~\cite{cbir_08_survey}. From the user perspective, \toolname{} is an interactive, multi-modal CBIR system that allows users to find relevant images from a large personal collection. In particular, \toolname{} is \emph{multi-modal} in  that the user provides a combination of natural language and example images, and it is \emph{interactive} in that the user can refine the query results by providing feedback through the \toolname{} interface.   From the system perspective, many prior CBIR tools 
search for target images using  metadata (e.g., where or  when an image was taken)~\cite{google_photos, apple_photos, piktures_app} or based on features extracted through machine learning techniques (e.g., lighting conditions and position of an object)~\cite{clothing_13}. In contrast, the backend underlying \toolname{} is based on neuro-symbolic program synthesis --- that is, it leverages the user's examples and natural language query  to synthesize a \emph{logical search query} utilizing pre-trained neural networks for object detection and classification.
In the remainder of this section, we focus on prior work that is closely  related to \toolname{} and refer the interested reader to existing surveys~\cite{cbir_00_survey,cbir_08_survey,cbir1,cbir2} for a more comprehensive overview of CBIR.

\paragraph{Expressing User Intent in CBIR} A key challenge in image retrieval is the \emph{intention gap}: the difficulty users face in articulating their task through queries~\cite{cbir_00_survey}.  Prior work aims to address this concern through  different modalities of input~\cite{sketch_matching_05,mental_image_05,simplicity,concept_map_10,paragraph_queries_15} and  multiple rounds of user interaction~\cite{whittlesearch, zhang2022tell, chen2020image, cosmo}. One line of work similar to \toolname{} is \emph{composed image retrieval}~\cite{Liu_2021_ICCV,Baldrati2023ComposedIR,Wen2023TargetGuidedCI}, which utilizes visual and textual modalities to \emph{jointly} specify the user’s intent.  In this line of work, an example image illustrates the concepts that the user is looking for, while the text query specifies what should be \emph{different}  (e.g. ``same dress but blue instead of red''). In contrast to such interfaces, \toolname{} uses natural language to \emph{directly} convey the user's intent rather than specifying what should be \emph{different} from a given image. In particular, users of \toolname{} utilize positive and negative images to \emph{clarify} ambiguities in the natural language query rather than providing them as a starting point for visual similarity search.

\paragraph{Relevance Feedback-Based Search Paradigms} Relevance feedback (RF) is a paradigm for interactively refining search results based on user feedback~\cite{ rf_03}. In many systems, users provide relevance feedback in the form of positive and negative images where positive examples correspond to those that are relevant to the user's query while negative examples are not~\cite{interaction_98,query_refinement_98,feedback_99,boosting_04,feedback_98,semantic_feedback_00,qcluster_03}.  \toolname{} is similar to these approaches in that the user can refine the initial query results by providing positive and negative examples. However, in contrast to many RF systems where examples are used to re-rank  the search results (e.g. ~\cite{query_refinement_98,feedback_98,qcluster_03}), \toolname{} uses positive and negative examples to extract \emph{hard semantic constraints} that the query results should or should not satisfy. 

\paragraph{Semantic Concepts for Images.} Another significant hurdle in image retrieval is the \emph{semantic gap}, which refers to the challenge of describing high-level semantic concepts using low-level visual features~\cite{cbir_00_survey}. Past research has explored deep learning techniques based on Convolutional Neural Networks (CNNs), including architectures like SqueezeNet ~\cite{squeezenet}, VGG ~\cite{vgg}, and ResNet ~\cite{resnet}, to address this problem. \toolname{} builds on recent advances in this field and leverages pre-trained neural networks for object detection and classification.  Prior work targets a variety of applications, including  geolocation ~\cite{cbirgeo1, cbirgeo2}, medical diagnosis ~\cite{cbirmed1, cbirmed2, cbirmed3, cbirmed4}, and interior decorating ~\cite{bell2015learning}. However, in contrast to most neural CBIR systems, \toolname{} learns new semantic concepts by composing existing neural networks via symbolic operators.

\subsection{Neuro-symbolic Programming for Images}

As mentioned earlier, \toolname{} performs image retrieval by synthesizing neuro-symbolic queries that combine pre-trained neural networks with symbolic operators. Specifically, \toolname{} first synthesizes a query  that is consistent with the user-provided input and then retrieves the desired images by executing the query on the user's dataset.  Hence,  \toolname{} is related to a line of recent work on neuro-symbolic programming for images~\cite{tian2018learning, pmlr-v97-young19a, mao2018the, pmlr-v119-huang20h, ellis18, Johnson_2017_ICCV, ReedF15, imageeye}.

\paragraph{General Visual-Reasoning Tasks.} 
Several recent works such as~\cite{visprog,vipergpt} have proposed using neuro-symbolic programming to automate image-related reasoning tasks, such as visual question answering (VQA), image editing, and object tagging. In particular, VisProg~\cite{visprog} proposes a neuro-symbolic DSL targeting images and uses in-context learning to synthesize programs in this DSL based on natural language. ViperGPT~\cite{vipergpt} proposes a custom Python API for visual reasoning tasks and synthesizes Python programs using this API, also based natural language queries. The back-end of \toolname{} synthesizes neuro-symbolic programs; however, it uses a combination of natural language queries and positive and negative examples. In particular, \toolname{} generates a so-called \emph{program sketch} by leveraging the natural language description and refines this sketch into a full query by utilizing the user-provided examples.

\paragraph{Specific Applications.} While VisProg~\cite{visprog} and ViperGPT~\cite{vipergpt} propose  general neuro-symbolic programming frameworks that can be adapted to several visual reasoning tasks, prior research has also developed more robust application-specific methods that use neuro-symbolic programming~\cite{tian2018learning, pmlr-v97-young19a, mao2018the, pmlr-v119-huang20h, ellis18, Johnson_2017_ICCV, ReedF15, imageeye}. Similar to our work, these efforts typically combine symbolic operators for higher-level reasoning with neural modules for perception, with the goal of learning new concepts in a few-shot manner. For example, Huang et al. ~\cite{pmlr-v119-huang20h} generate programmatic \emph{referring expressions} that identify specific objects in an image in terms of their attributes and relationships to other objects. This work focuses on locating a single object, whereas our DSL expresses image search tasks that involve multiple objects. In addition,  their focus is on a synthetic dataset with geometric shapes, while our focus is on more realistic images with faces, text, and arbitrary objects.

In the domain of image manipulation, ImageEye ~\cite{imageeye} allows users to automate batch image editing tasks using neuro-symbolic  programming. In particular, ImageEye captures demonstrations of a user editing an image and then synthesizes neuro-symbolic programs that are consistent with the demonstration. In contrast to \toolname{}, ImageEye does not utilize natural language; instead, it requires the user to demonstrate the task by applying actions to selected parts of an image.

Another related work in this space is RAPID~\cite{wang2023rapid}, which is a system for automated image labeling. The idea behind RAPID is to express new visual concepts (e.g., \emph{chef}) as logical combinations of existing concepts and then learn these concept definitions from positive and negative examples.    For instance, RAPID may learn that an image should be labeled  ``chef'' if there is food or a bowl in the image. In contrast to \toolname{}, RAPID does not utilize natural language descriptions, and thus lacks the inductive bias for efficient image search.  Additionally, RAPID uses a different learning approach based on first-order inductive logic learning.

\section{Usage Scenario}\label{sec:usage}

\begin{figure}
\centering
\begin{minipage}[t]{0.75\textwidth}
    \centering
    \includegraphics[scale=0.19]{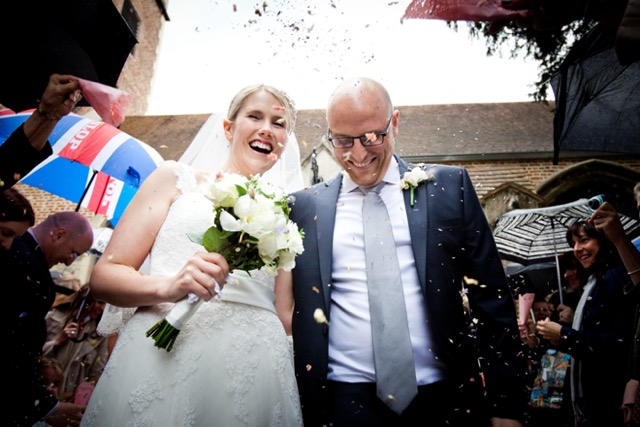}
    \includegraphics[scale=0.19]{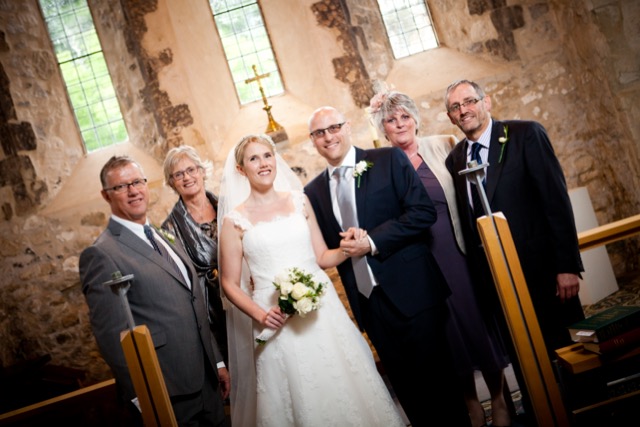}
    \includegraphics[scale=0.14]{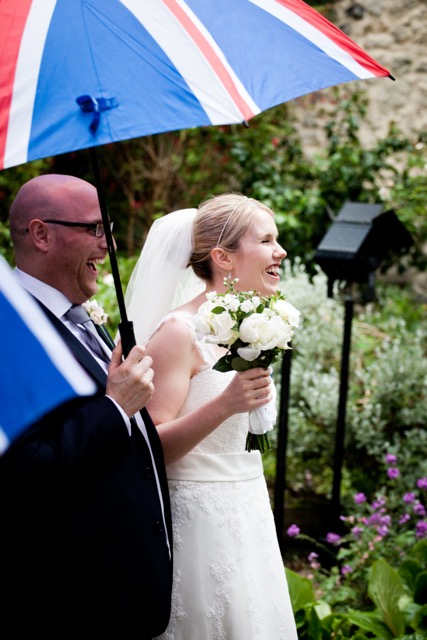}
    \label{fig:usage_images}
\end{minipage}
\unskip\ \vrule\
\begin{minipage}[t]{0.21\textwidth}
    \centering
    \includegraphics[scale=0.21]{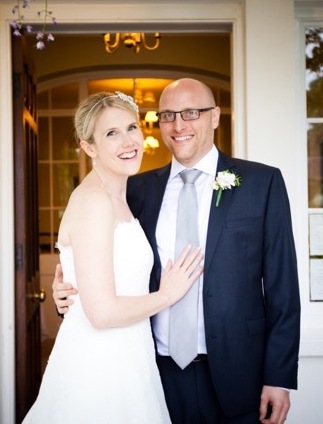}
    \label{fig:negative_image}
\end{minipage}
\caption{Left: Three images that matches John's intent: the bride and groom are next to each other, with the bride holding flowers. Right: an image that is incorrect image because the bride is \emph{not} holding flowers.} 
\Description{Left: three images from a wedding where the bride and groom are next to each other, and the bride is holding flowers. Right: an image from the wedding where the bride and groom are standing next to each other, but the bride is not holding flowers. The groom has flowers pinned to his jacket.}
\label{fig:usage_and_negative_images}
\end{figure}

This section illustrates the interface and features of \toolname{} through a use case inspired by real-world scenarios described in online blogs ~\cite{weddingblog}. In this example, a  photographer, John, is preparing a wedding photo album and needs to locate specific images among hundreds of photos he took during the wedding. As part of this process, John needs to find  photos in which the bride, Alice, and the groom, Bob, are next to each other and where Alice is holding flowers. For example, the first three images in Figure~\ref{fig:usage_and_negative_images}   meet John's requirements but the last one does not. 
John finds this task challenging to perform using existing similarity-based search tools, as there are a lot of other images containing Alice, Bob, and flowers, but many of these images do not match his \emph{logical constraints} --- for example, there is another person between Alice and Bob or Alice is not holding flowers. We now illustrate how John can use \toolname{} to perform this task and avoid significant manual labor. 

\begin{figure}
    \centering
    \includegraphics[scale=0.26, trim=20 0 300 100, clip]{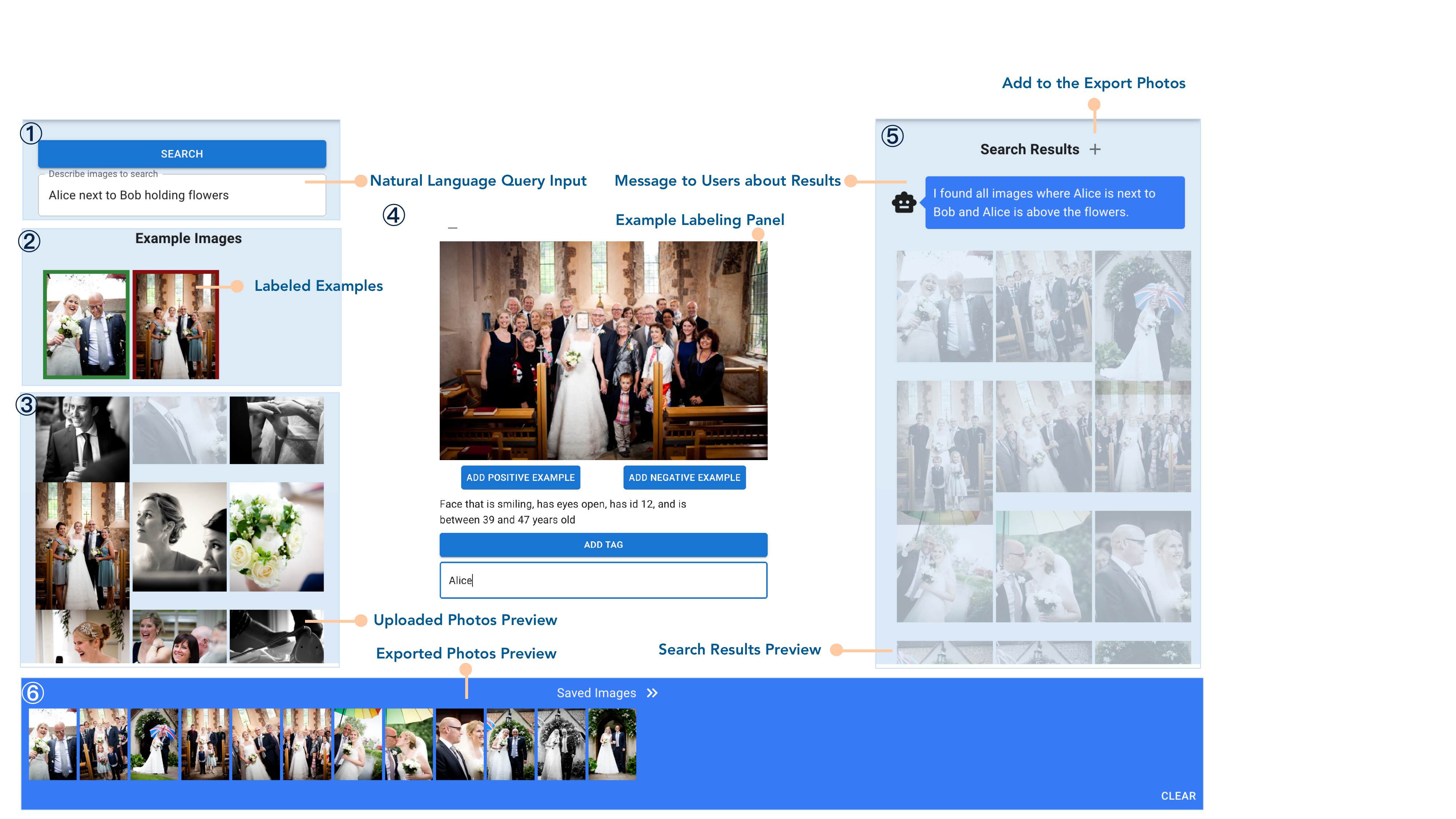}
    \caption{The \toolname{} interface has six main panels: (1) The user enters a natural language query describing the images to be searched. (2)  The example images panel highlights all the positive and negative images that the user has already labeled. Positive examples are wrapped in a green box and negative examples are wrapped in a red box. (3) The album preview panel displays all the photos in the album to be searched from. (4) Once the user selects a photo to label, the example labeling panel displays the image and the example labeling buttons. (5) The search results panel shows all the images that \toolname{} finds that match both the natural language description and the labeled examples, along with a natural language explanation. (6) The photo export panel shows all the images selected by the user as the final search results.}
    \label{fig:tool_interface}
    \Description{Web page with three columns. In the left column, there is a text box and a "Search" button where the user can submit a natural language query. Below that, there are rows of images that the user has selected as examples. Below that, there are rows of images from the user's album. In the center column, there is a single image displayed. In the image, the bride's face is highlighted. Below the image are buttons that say "Add Positive Example" and "Add Negative Example", a button that says "Add Tag", and a natural language description of the bride's face. In the right column, there is a header that says "Search Results", a + button for exporting the search results, and rows of images in the search results. At the bottom of the web page, there is a horizontal list of exported images.}
\end{figure}

\begin{figure}
    \centering
    \begin{minipage}[c]{0.63\textwidth}
      \includegraphics[scale=0.26, trim=20 0 800 400, clip]{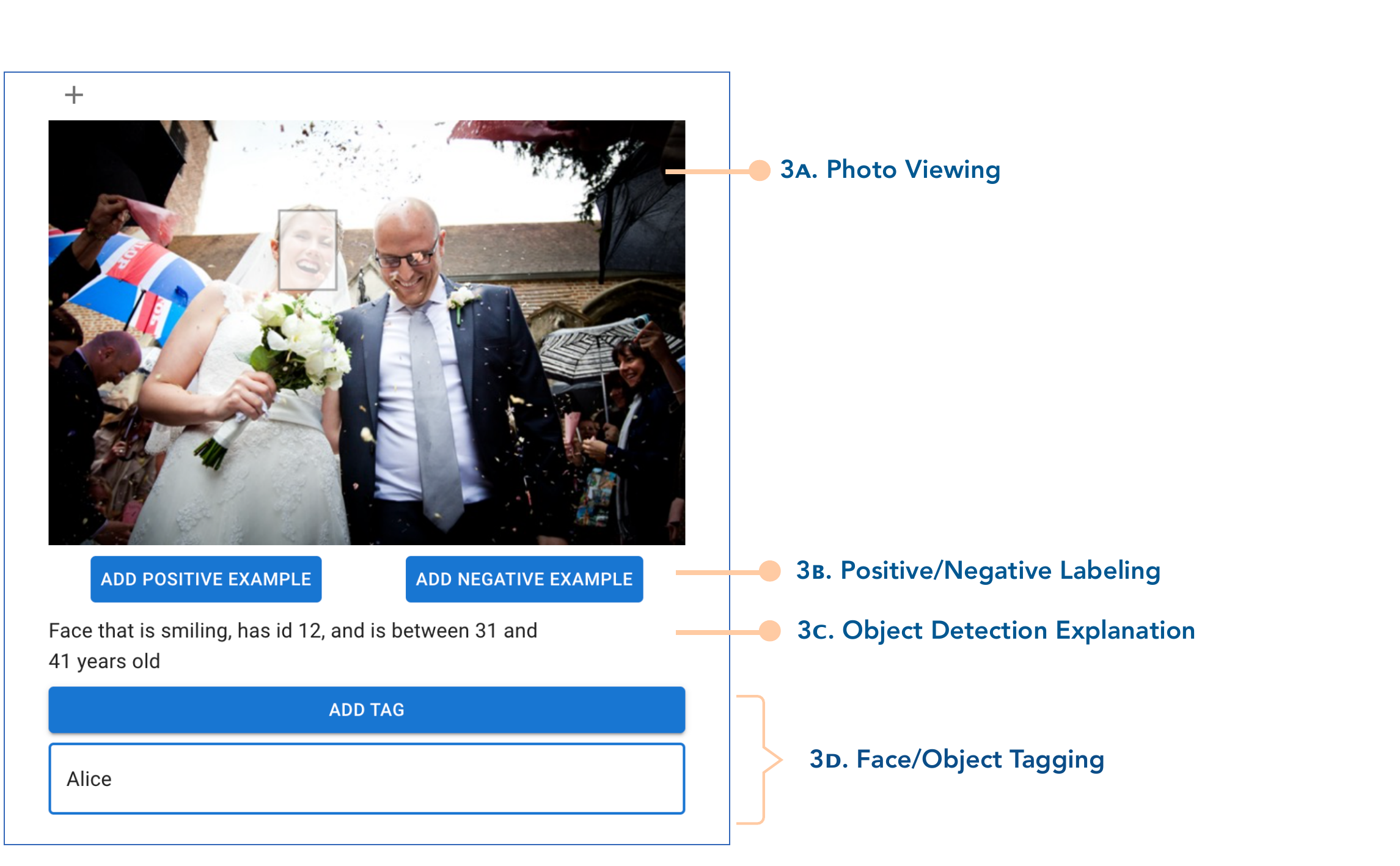}
  \end{minipage}\hfill
  \begin{minipage}[c]{0.36\textwidth}
 \caption{The example labeling panel consists of 4 elements. (a) A view of the photo to be labeled. When a user hovers over it, each object identified by the detector is highlighted with a square box, with the detailed description of the detected object shown in (c). (b) asks the user to label the photo either as a positive or negative example. (d) is a tagging interface so the user can give semantic meanings to the detected face or object. In this particular example, the user is tagging the bride with the name ``Alice'' so that they can refer to the bride in the query. }
    \label{fig:example_labeling}
    \Description{A closeup of the center column of the web page from Figure ~\ref{fig:tool_interface}. There is an image from the wedding displayed where the bride's face is highlighted. Below the image there are buttons that say "Add Positive Example" and "Add Negative Example", a natural language description of the bride's face, a button that says "Add Tag", and a text box with "Alice" typed into it.}
  \end{minipage}

\end{figure}

\autoref{fig:tool_interface} shows the general interface of \toolname{}, which contains three main components: (1) a task specification panel (\autoref{fig:tool_interface}--\circled{1} to \circled{4}) that allows  the user to communicate their intent using a combination of natural language queries and image labels, (2) a search result panel (\autoref{fig:tool_interface}--\circled{5}) that shows the results from the current search query, and (3) a saved images panel (\autoref{fig:tool_interface}--\circled{6}) for saving and exporting the desired images.
Using this interface,  John can complete his task by performing the following steps:
\begin{enumerate}
    \item \emph{Load images.} John first loads all the images to \toolname{} and then sees the interface shown in Figure~\ref{fig:tool_interface}. 
    \item \emph{Write natural language query.} The user interface exposes a search box where the user can type a natural language query (Figure~\ref{fig:tool_interface}--\circled{1}) and a panel displaying  thumbnails for all uploaded photos  (Figure~\ref{fig:tool_interface}--\circled{3}). In a typical use case, the user  starts by entering a natural language query, such as ``Alice next to Bob holding flowers'', and clicks the ``Search'' button. 
    \item \emph{Tag objects.} In this example, \toolname{} does not yet know who Alice and Bob are, so, in the search results panel (Figure~\ref{fig:tool_interface}--\circled{5}), \toolname{} displays a message communicating this missing information. John resolves this ambiguity by selecting an image and tagging Alice and Bob's face in the labeling panel (Figure~\ref{fig:tool_interface}--\circled{4}). Figure~\ref{fig:example_labeling} provides a more detailed view of the labeling panel. When John selects a photo, \toolname{} shows the full-size photo in the center of the labeling panel. The photo is annotated with object detection and classification results to help the user understand what the underlying computer vision tools ``see'' in that image. For example, when the user hovers over a part of the photo, \toolname{} displays detected objects as a square box, as shown in Figure~\ref{fig:example_labeling}. Additionally, the interface displays a natural language description of the classification results for that object. For example, Alice's face in Figure~\ref{fig:example_labeling} is further categorized as smiling and between 31 to 41 years old. In this scenario,  John clicks on the face of the bride and labels the face as  Alice (see 3d Figure~\ref{fig:example_labeling}). At this point, \toolname{} learns to associate this face with Alice, ensuring that  she can be  referenced in future search queries without additional user interaction. John uses the same panel to similarly detect the groom's face and label it as Bob.
    \item \emph{Select positive examples.} After tagging these faces, John clicks ``Search'' to see the updated results. This time, \toolname{} is not sure about the concept of ``holding flowers''  and asks John to illustrate this concept by providing examples.  John labels the first image in Figure~\ref{fig:usage_and_negative_images} as a positive example using the labeling panel  (Figure~\ref{fig:tool_interface}--\circled{4}) and clicks ``Search'' again.
    \item \emph{Select negative examples.} This time, instead of asking for clarification, \toolname{} shows all relevant images in the result panel (Figure~\ref{fig:tool_interface}--\circled{5}), along with an natural language explanation of how it generated these results. After looking at the explanation and inspecting the results, John notices that the results contain all relevant photos but also some extra ones, specifically those where there are flowers, but Alice is not holding them (e.g., the last photo in Figure~\ref{fig:usage_and_negative_images}, where Bob's boutonnière is visible). To further refine the search results, John labels this photo as a negative example and does another round of search. This time, \toolname{} returns all photos of Alice next to Bob with Alice holding flowers. As a final step, John clicks on the ``+'' sign located at the top of the search results section (Figure~\ref{fig:tool_interface}--\circled{5}) and 
all the added photos are displayed in Figure~\ref{fig:tool_interface}--\circled{6}.
\item \emph{Manually add/remove images.}   Upon inspection, John finds that there is one photo in the results in which Alice's flowers are sitting in front of her on a table, but she is not holding them. To exclude that photo from the search results, John selects the photo from Figure~\ref{fig:tool_interface}--\circled{6} and clicks the ``-'' button on the top left of Figure~\ref{fig:tool_interface}--\circled{4} to remove this image from the export results. Once John is happy about the results, he clicks the ``>>'' button in Figure~\ref{fig:tool_interface}--\circled{6} to export the results to a user-defined directory.
\end{enumerate}

In summary, John is able to find all the photos he wants to retrieve by first providing a natural language query and then  iteratively refining this query by tagging objects  and labeling photos as positive or negative examples. In this process, he benefits from the following design decisions behind \toolname{}:

\begin{itemize}[leftmargin=*]
\item {\bf Multimodal Inputs.} \toolname{} grants John the versatility to articulate his search criteria both using natural language prompts and positive and negative examples. On one hand, solely relying on natural language introduces several potential ambiguities: For example, who are Alice and Bob, and who should be holding the flowers? On the other hand, solely relying on examples would be quite cumbersome, as John would need to provide several more examples to convey his intent. In contrast, the \emph{combination} of natural language and image annotations allows John to succinctly and efficiently convey his  intent.
    
\item {\bf Semantic search. } In our example, John's search query is quite specific: First, Alice and Bob must be next to each other, and, second, Alice should be holding flowers. Such search queries are out of scope for existing image retrieval systems, as they cannot reason about \emph{relationships} between objects within an image.  In contrast, our approach performs search by first synthesizing a \emph{neuro-symbolic program} and then executing that program on all images. This synthesis-based approach allows \toolname{} to perform structured image search tasks where the goal is to find images that conform to non-trivial logical constraints.

 
 \item {\bf Feedback-guided refinement.} Rather than presuming John to deliver flawless instructions from the outset, \toolname{} employs an interactive feedback mechanism. As illustrated in our example, \toolname{} recognizes ambiguous elements in John's description and proactively seeks clarification via natural language prompts. 

\item {\bf Fast synthesis procedure.} To ensure that John does not have to wait a long time when interacting with the tool, \toolname{} adopts an efficient synthesis approach to find useful programs. Each synthesis run takes between 0.36 and 4.8 seconds, making it feasible to use \toolname{} in an interactive fashion. 

\end{itemize}


\section{System Architecture and Implementation}\label{sec:impl}

In this section, we discuss the design and implementation of \toolname{}. As mentioned earlier, \toolname{} performs image search by first  synthesizing a program in a neuro-symbolic  domain-specific language (DSL) and then applying that program to all images in a  collection.  In this section, we first provide an overview of the image search DSL and then explain the internal workflow of \toolname{} in more detail. 

\subsection{Image Search DSL}

\begin{figure}
     \centering
     \begin{minipage}[b]{0.3\textwidth}
         \centering
         \includegraphics[width=1.4in]{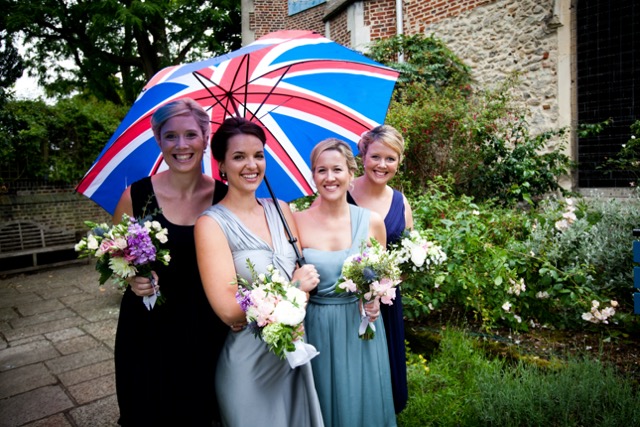}
         \caption{An example image.}
         \Description{An image from the wedding dataset. Four bridesmaids are standing together. Each of them holds a bouquet of flowers.}
         \label{fig:example1}
     \end{minipage}
     \begin{minipage}[b]{0.68\textwidth}
         \centering
    \[
    \begin{array}{rcl}
        E & := &  r(t_1, \ldots, t_n) \\
        & | & \ E \rightarrow E  \ | \ E \wedge E \ | \ E \vee E \ | \ \neg E \ |  \ \exists x. E \ | \ \forall x. E \\

        r & := & \textsf{HasType} \ | \ \textsf{HasEmotion} \\
          & | & \textsf{HasRelation} \ | \ \textsf{HasProperty}  \\ 
                t & := & x \ | \ c \\
    \end{array}
    \]
         \caption{Image Search DSL. All predicates are binary except for HasRelation (ternary).}
         \label{fig:dsl}
     \end{minipage}
\end{figure}

\toolname{}'s DSL, shown in Figure~\ref{fig:dsl}, is designed to express a wide array of  image search tasks. At a high level, a program in this DSL is similar to a first-order logic formula, and evaluates to either true or false given an input image. 

The primitives in this DSL are predicates of the form $r(t_1, \ldots, t_n)$ where $r$ is an n-ary relation and each $t_i$ is a term (constant or variable). The \toolname{} DSL contains many built-in predicates such as the binary relations \textsf{HasEmotion}($t, c$) and \textsf{HasType}($t, c$),  as well as ternary relations  such as $\textsf{HasRelation}(t_1, t_2, c)$. Note that the semantics of the predicates are determined using neural models; hence, we refer to this DSL as \emph{neuro-symbolic}. For example, $\textsf{HasType}(o, \textsf{Car})$ is determined by using an object classification model to check whether $o$ is classified as a car. 
Similarly, the truth value of $\textsf{HasRelation}(o_1, o_2, \textsf{Above})$ can be determined by using an object classification model that identifies bounding boxes around objects $o_1, o_2$ and then using the resulting coordinates to check whether $o_1$ is above $o_2$. As standard in first-order logic, predicates can be combined using boolean connectives. Additionally, our DSL allows quantification over variables to test whether an image contains \emph{some} object with a given property (requiring existential quantification) or whether \emph{all} objects have a certain property (requiring universal quantification). 

In our implementation, the truth value of atomic predicates is determined using the Amazon Rekognition library. The pre-trained neural nets supported by this library can detect and locate a wide array of objects in image. In particular, this library can be used to identify bounding boxes for different objects in the image, and to determine their types (e.g., cat, car, person etc).  Rekognition can also detect properties of human faces (e.g. whether a face is smiling or has open eyes) and identify the same face across multiple images. Overall, it is this \emph{combination} of logical operators and neural models that allows \toolname{} to express a rich class of structured image search tasks.  



\begin{example}
    Consider the following program:
    \[ \forall x.  ( \textsf{HasType}(x, \textsf{Face}) \rightarrow (\textsf{HasProperty}(x, \textsf{Smiling}) \land \exists y. (\textsf{HasType}(y, \textsf{Flower}) \wedge \textsf{HasRelation}(x, y, \textsf{Above}))))\]
In this program, the universal quantifier $\forall x$ indicates that every object $x$ identified in the image must obey the subsequent condition. In particular, if $x$ is identified to be a human face, then $x$ must be smiling, \emph{and} there must exist an object $y$ in the image such that $y$ is identified to be a flower, where $x$ is above $y$. Put simply, this program can be used to find images where every person is smiling and holding flowers, as in Figure \ref{fig:example1}. Note that the concept of ``$x$ holds $y$'' is approximated by checking a spatial relationship between $x$ and $y$. 
\end{example}




\subsection{PhotoScout Synthesizer}

\begin{figure}
\vspace{20px}
\includegraphics[scale=.4, trim=0cm 17cm 30cm 2cm]{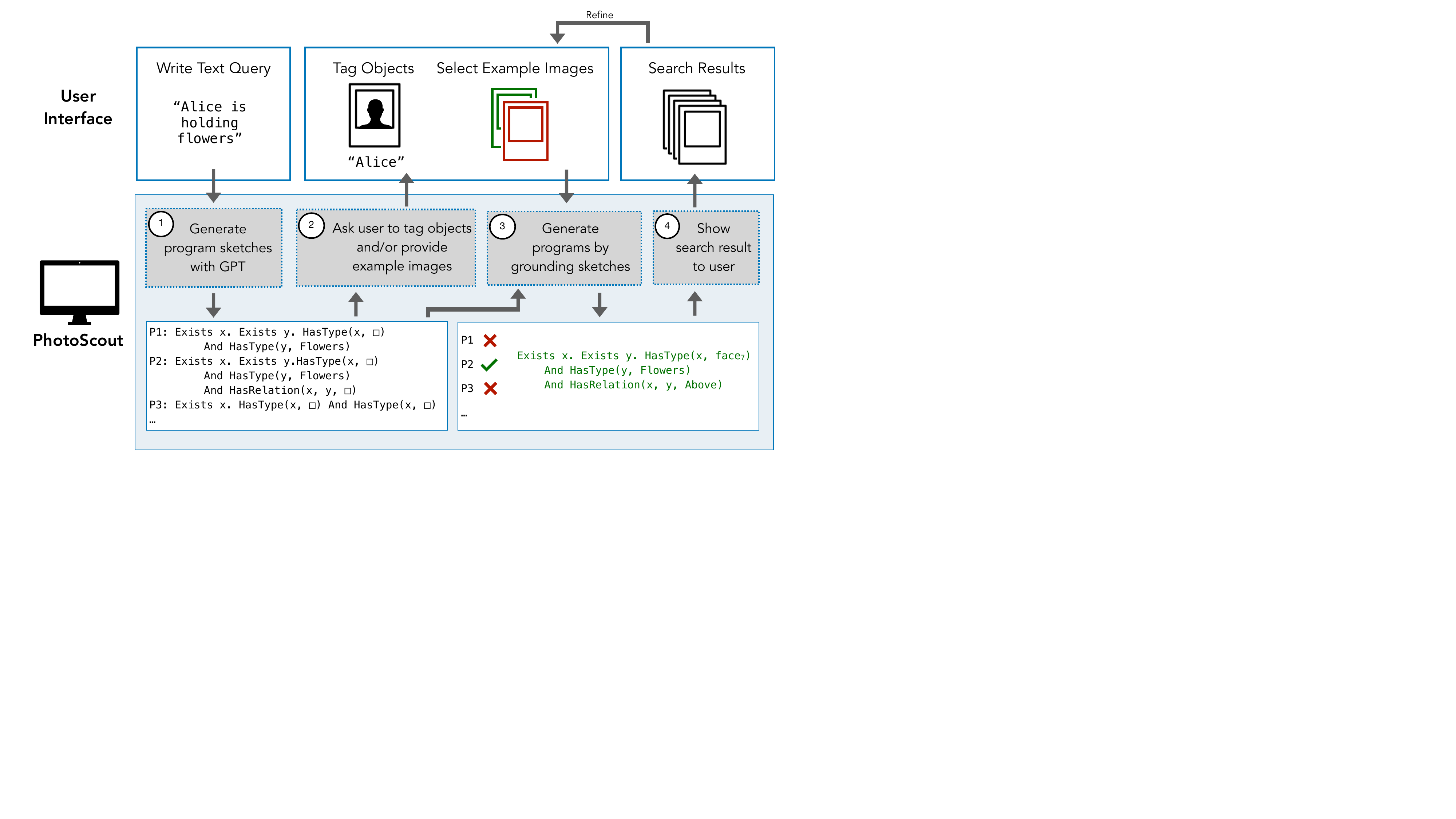}
\caption{The architecture of the \toolname{} system.}
    \label{fig:arch}
    \Description{Two components: user interface and \toolname{} backend. In the user interface, the user writes a text query which is passed to \toolname{}. \toolname{} generates program sketches with GPT, and asks the user to tag images and/or provide example images. The user tags objects and selects example images, and \toolname{} uses these additional inputs to ground the program sketches. \toolname{} selects a complete program and shows the search result to the user. The user can refine their results by adding additional example images. }
\end{figure}

In this section, we describe \toolname{}'s underlying synthesis engine, which is depicted schematically in Figure~\ref{fig:arch}. Given the initial natural language query, \toolname{} starts by generating  a \emph{program sketch} containing holes (i.e., unknowns denoted as $\square$).  Intuitively, \toolname{} cannot directly generate a program from the natural language query because some of the concepts used in the NL description have to be grounded. For example, given a query like ``Alice is holding flowers,'' the synthesizer has no idea what Alice corresponds to or how the concept of ``holding'' can be implemented in our DSL. To  instantiate the program sketch into a complete program, \toolname{} interacts with the user by asking them to tag objects or provide examples (Step 2 in  Figure~\ref{fig:arch}). In the third step, the synthesizer fills the holes in the sketch by performing enumerative search over the space of sketch completions and discarding those programs that do not satisfy the examples. In the  final step, the synthesized program is applied to all uploaded images and displayed to the user. If the search results are unsatisfactory, the user can refine the query by providing more positive and negative examples. We now explain each of the steps in this process in more detail.

\begin{figure}
    \centering
    \includegraphics[scale=.35, trim=0cm 27cm 28cm 0cm]{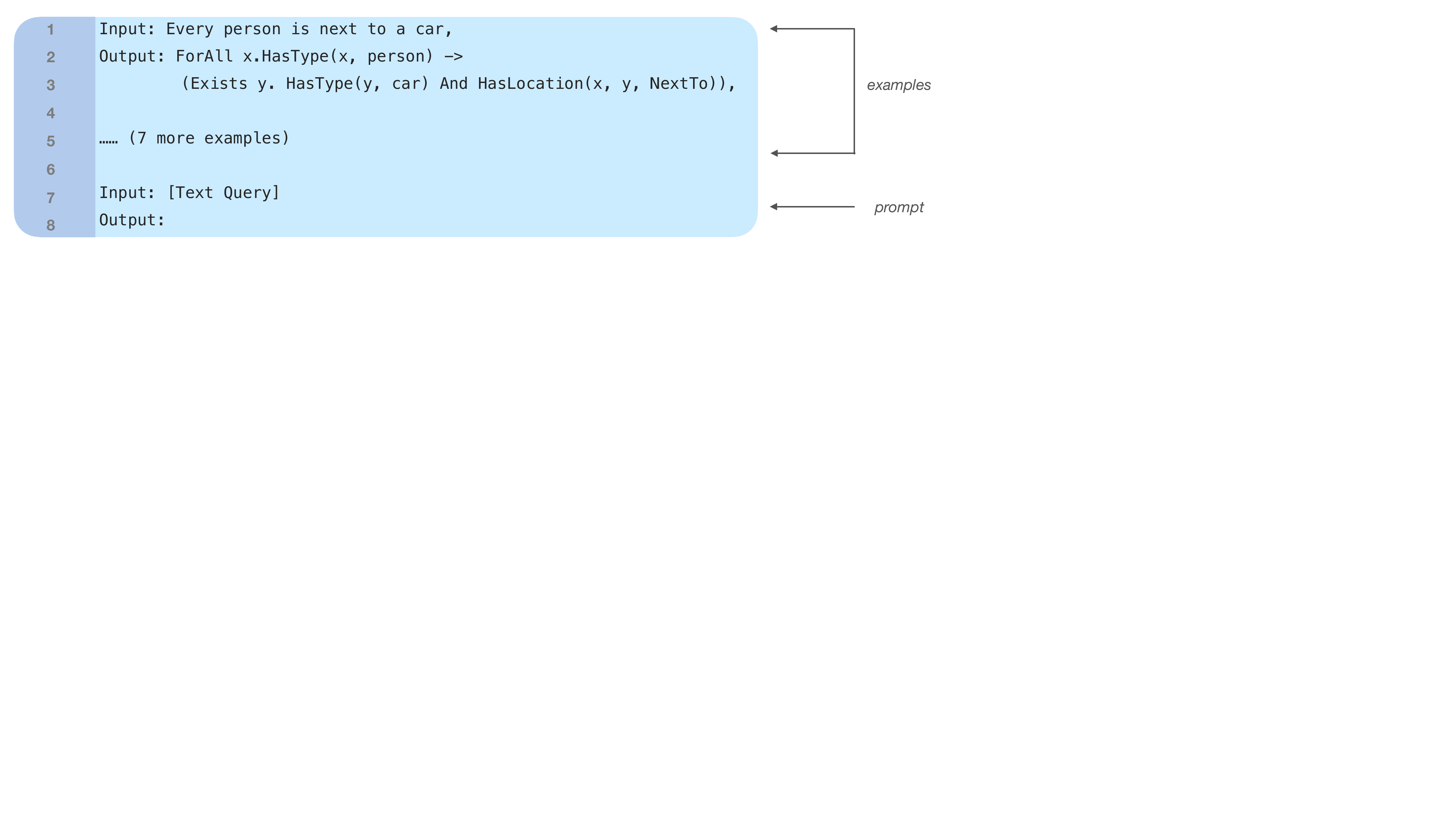}
    \caption{GPT prompt for generating program sketches from a user's text query.}
    \Description{An example of one of the (query, program) pairs given to GPT. The full prompt is in the Supplemental Materials.}
    \label{fig:gpt-query}
\end{figure}

\paragraph{Step 1: Generate program sketches.} Motivated by the success of few-shot prompting  in similar domains~\cite{brown2020language, zhuo2023robustness, chen2023data}, \toolname{} obtains program sketches by prompting  GPT-3.5 Turbo.\footnote{While a different LLM could be used for the purpose of sketch generation, we use GPT in our implementation because we found it to more effective than alternative models that we tried.} The key idea is to provide GPT with examples of representative natural language and  program pairs and then ask it to generate a program for the user's NL query. Figure \ref{fig:gpt-query} shows an example of such a prompt where we provide the LLM with a manually curated set of representative (query, program) pairs as well as the natural language query of interest.. \toolname{} asks GPT to generate 20 answers  to this prompt in order to increase the likelihood that \emph{one} of the results match the user's intention. For each result returned by GPT, \toolname{} attempts to parse the string into a program in our DSL. If, during parsing, \toolname{} encounters a predicate or constant that it does not recognize, it  replaces that construct with a hole $\square$. If parsing fails for a different reason, that program sketch is discarded. 

\begin{example}
    Given the text query \texttt{``Alice is holding flowers''}, GPT may generate the program 
    \begin{align}
        \exists x. \exists y. \ (\textsf{HasType}(x, \textsf{Alice}) \wedge \textsf{HasType}(y, \textsf{Flowers}) \wedge \textsf{HasRelation}(x, y, \textsf{Holding}))
    \end{align}
    However,  since \textsf{Alice} is not an object category recognized by the object detector and \textsf{Holding} is not a predicate in the DSL's grammar, these constructs will be replaced with holes. Thus, the following program sketch will be produced:
    \begin{align}
    \exists x. \exists y.\ (\textsf{HasType}(x, \square_1) \wedge \textsf{HasType}(y, \textsf{Flowers}) \wedge \textsf{HasRelation}(x, y, \square_2))
    \end{align} 
\end{example}

\paragraph{Step 2: Query the user.} If the program generated in Step 1 contains holes,  \toolname{}  prompts the user to provide additional information by (1) tagging objects that are not recognized by the object detector and (2) adding example images that clarify the meaning of unknown predicates. 
Tags and examples images both allow the user to clarify the meaning of their natural language query, but in complementary ways. Tags allow grounding unknown terms like people's names in the user's NL query, whereas positive and negative examples allow the synthesizer to learn logical constraints and concepts (e.g., the concept of ``holding'') in a data efficient way.

\begin{example}
    Given the  program sketch
    \begin{align}
    \exists x. \exists y. (\textsf{HasType}(x, \square_1) \wedge \textsf{HasType}(y, \textsf{Flowers}) \wedge \textsf{HasRelation}(x, y, \square_2))
    \end{align}
    where $\square_1$ and $\square_2$ were derived from $\textsf{Alice}$ and $\textsf{Holding}$, respectively, \toolname{} will display the following message to the user: 
    \texttt{``I don't know the terms `Alice' and `Holding'. Can you provide a few positive and negative examples and/or tags to show me what you mean?'' } The user can easily ground the name ``Alice'' by using the \toolname{} interface to add a tag. However, the concept of ``holding'' is harder to explain through a tagging, as it corresponds to a binary relationship between two objects. In this case, the user can help \toolname{} learn this concept by providing a few positive examples where Alice is holding the flower and a few negative examples of those where there is a flower in the picture but Alice is \emph{not} holding them.
\end{example}

\paragraph{Step 3: Sketch completion.} While object tagging helps resolve many sources of ambiguity in the natural language, \toolname{} needs to perform enumerative search over possible sketch completions to find a program that is consistent with all positive and negative examples. Given a program sketch $P$ and a a set of  positive and negative examples $\mathcal{E}^+ \cup \mathcal{E}^-$, \toolname{} enumerates possible completions of $P$ by instantiating each hole with  a constant and then evaluating the resulting query $Q$ on each example in $\mathcal{E}^+$ and $\mathcal{E}^-$. If $Q$ evaluates to true (resp. false) for all examples in $\mathcal{E}^+$  (resp. $\mathcal{E}^-$), then $Q$ is retained as a viable completion of the sketch. 
 Among all programs that are consistent with the examples, \toolname{} chooses the simplest program, where simplicity is defined in terms of the number of nodes in the program's abstract syntax tree. Note that enumerative search is tractable in this context because each sketch contains no more than a few holes.




\begin{example}
Consider the following partial program:
\begin{align}
    \exists x. \exists y. \ (\textsf{HasType}(x, \textsf{face}_{\textsf{n}}) \wedge \textsf{HasType}(y, \textsf{Flowers}) \wedge \textsf{HasRelation}(x, y, \square))
\end{align}
Suppose that the user has added the first three images in Figure \ref{fig:usage_and_negative_images} as positive examples and the last one as negative.  Consider the completion $P'$ of this partial program where $\square$ has been filled with \textsf{NextTo}. For each positive example image $\mathcal{I}^+$, $P'(\mathcal{I}^+)$ is true, as Alice is adjacent to flowers in each of these images. However, for the negative example image $\mathcal{I}^+$, $P'(\mathcal{I}^-)$ is also true. Thus, $P'$ is not a valid completion of the program. However, the completion $P''$ where $\square$ has been filled with \textsf{Above} is a valid completion, as Alice's face is below flowers in every positive example, but not in the negative example. 

\end{example}

Note that this step is useful even if none of the program sketches contain holes, as example images will filter out complete programs that do not match the user's intent. 

\paragraph{Step 4: Display search results} The last step in the process is to execute the synthesized program $P$ on all input images and display those images $I$ for which $P(I)$ yields true. Since the number of search results may be quite large, \toolname{} also generates a natural language explanation of what the program does. Such explanations are intended to help users quickly uncover unintended behaviors without having to look through a large set of images and inspecting each one. \toolname{} generates these NL descriptions through few-shot prompting of an LLM: In particular, given a few examples of programs and their corresponding NL description, \toolname{} prompts GPT to produce an natural language description of the programmatic query. 

\begin{example}
Suppose that $P$ is the program 
\begin{align}
    \exists x. \exists y.\textsf{HasType}(x, \textsf{Alice}) \wedge \textsf{HasType}(y, \textsf{Flowers}) \wedge \textsf{HasRelation}(x, y, \textsf{Above}).
\end{align}
Then GPT may generate the following natural language explanation: \texttt{``I have found all images that contain Alice and  flowers and where Alice's face is directly above the flowers.''}
\end{example}

Even after \toolname{} generates a correct program, it may not produce \emph{exactly} the desired output for two main reasons: First, some concepts such as ``holding'' may not be perfectly expressible in our DSL. For instance, in our running example, we  approximate the concept of holding through a coarse spatial relationship between objects (e.g., if face $x$ is directly above object $y$, then $x$ is holding $y$). Second, even when all concepts are perfectly expressible, the program may not produce the desired output due to inaccuracies in the underlying neural model. For instance, if the face recognition model does not correctly classify Alice's face, then a photo containing Alice may not appear in the search results even though it should. \toolname{} deals with this problem by allowing users to manually add or remove images through the Saved Images panel of the user interface.

\subsection{Design Considerations}

We conclude this section by summarizing and justifying some of the design considerations underlying \toolname.

\paragraph{User Interface} The design principles of \toolname{}'s user interface reflect the requirements of structured image search tasks. As seen in the usage scenario, a structured search task may be simple and intuitive to describe in natural language, but contain ambiguities that are easier to resolve through visual examples. Hence, our interface allows the user to interactively refine the search results. In a typical workflow, the user begins their search by writing a natural language query, which may contain unknown terms and concepts that need to be \emph{grounded}. 
To help the user understand which terms  need to be grounded via user interaction, \toolname{} generates natural language explanations of what it does \emph{not} understand. The user then can then interact with \toolname{} to teach it new concepts. In particular, constants such as people's names are natural to teach via object tagging, whereas new predicates (e.g., ``holding'') can be demonstrated using positive and negative examples.  Furthermore, the user can provide these examples in a piecemeal fashion by providing one example at a time, re-running the synthesizer, and inspecting the search results.


\paragraph{System Implementation} Recall that \toolname{}'s system represents search tasks as programs in a DSL. Utilizing DSLs for visual tasks is an approach established in prior work \cite{imageeye, vipergpt, visprog, progprompt}. In the context of structured image retrieval, we believe that such a DSL-based approach is a particularly good fit, as the user wishes to find images that satisfy certain logical constraints. 

We note that any DSL imposes a tradeoff between \emph{expressiveness} and \emph{reliability}: the more expressive the DSL, the larger the space of tasks it can represent, leading to a harder  \emph{synthesis} problem. On the other hand, if the DSL is too restrictive, it may not be able to express image search queries that arise in practice. 
Our DSL maintains a balance between these two properties, capturing a wide array of structured image search tasks while keeping a compact structure similar to first-order logic that facilitates synthesis.

\toolname{} generates image search queries by using an LLM to ``translate'' the user's NL description to a program sketch in the DSL. This approach allows the synthesizer to extract as much information as possible from a coarse search query expressed in natural language. However, because the user's description may be ambiguous or contain new concepts that are not captured via pre-trained neural models, \toolname{} grounds new concepts via user interaction, which takes two forms: Object tagging allows the user to conveniently ground names, whereas positive and negative examples allow grounding unknown relations and resolving ambiguities.

\section{User Study}\label{sec:user_study}

To understand how people interact with \toolname{} and  gain deeper insight about the strengths and limitations of the proposed interface, we conducted a within-subjects evaluation centered on the following questions:
\begin{enumerate}[leftmargin=*]
    \item Does \toolname{} improve user efficiency and accuracy compared to a baseline{} image search tool?
    \item Does the proposed multi-modal interface help users express their intent?
    \item Are users more confident about the accuracy of the search results compared to the baseline{} tool?
    \item What strategies do people adopt when interacting with \toolname{}?
\end{enumerate}

In the remainder of this section, we first describe the baseline{} tool and our user study procedure. Afterwards, we   present both quantitative and qualitative analyses of the user study results.

\subsection{Baseline Tool: \baseline{}}

As a basis of comparison, we implemented a graphical user interface around OpenAI's CLIP model, which is a state-of-the-art neural network for learning visual concepts from natural language supervision. Given a dataset $\mathcal{D}$ of images and a query (in the form of text or image), the CLIP model assigns a score  to each image in $\mathcal{D}$ that reflects its similarity to the  given query. 

Our baseline tool, henceforth called \baseline{}, is a wrapper around this CLIP model. Specifically, \baseline{}\ implements a graphical user interface that allows users to input a query on a set of uploaded images. The query can either be a text description of the search task or a photograph that exemplifies the target search results. \baseline{} simply queries the CLIP model and returns all images in the dataset whose score exceeds some threshold.  The \baseline{} interface allows users to further refine the search results by manually adding or removing images to and from the result returned by the CLIP model. 

\begin{figure}
    \centering
\includegraphics[scale=0.3, trim=0 0 25cm 15cm,clip]{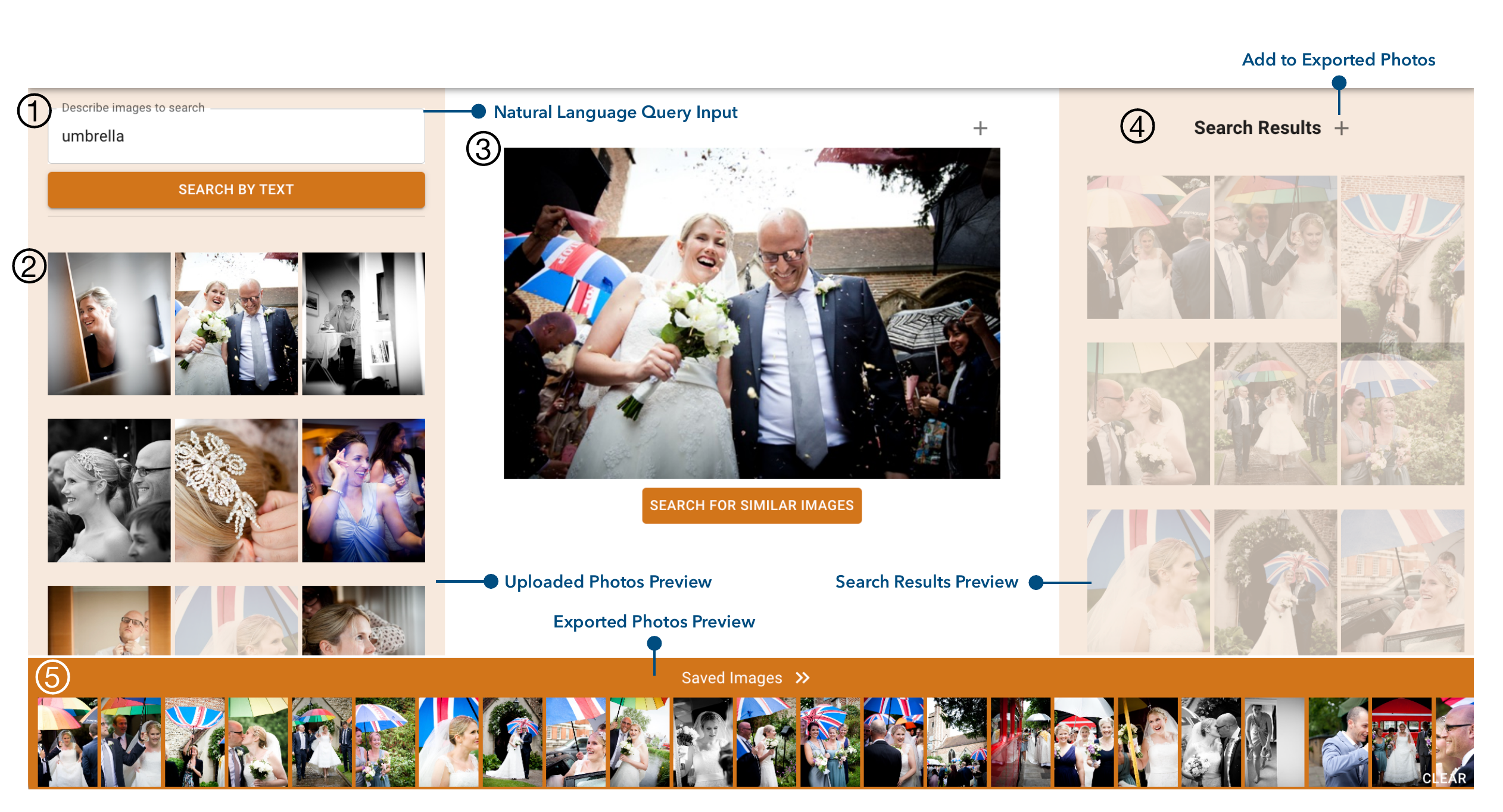}
    \caption{The \baseline{} interface has five main panels: (1) The user enters a natural language query describing the images to be searched. (2) The album preview panel displays all photos in the target album. (3) The photo view panel displays an image and allows users to search for similar images. (4) The search results panel shows all images that \baseline{} finds that match the user's query. (5) The photo export panel shows all images selected by the user as the final search result.}
    \label{fig:enter-label}
    \Description{Web page with three columns. In the left column, there is a text box and a "Search" button where the user can submit a natural language query. There are rows of images from the user's album. In the center column, there is a single image displayed. Below the image is a button that say "Search for Similar Images". In the right column, there is a header that says "Search Results", a + button for exporting the search results, and rows of images in the search results. At the bottom of the web page, there is a horizontal list of exported images.}
\end{figure}

\baseline{} allows users to search for images that are similar to either a query image or an open-ended text query.  \baseline{} does not utilize any hard constraints and may return images that do not precisely match a user’s query. A central question of this user study is: does \baseline{} suffice for performing structured image retrieval tasks, or is a tool specially designed for such tasks necessary? Further, we explore the specific features of \baseline{} that make structured image search difficult, compared with \toolname{}. \baseline{}'s interface mirrors \toolname{} as closely as possible so as to reduce the number of confounding factors in our comparison.




\subsection{User Study Procedure}

We recruited a total of 25 participants for our user study. Among these participants, 23 are in the 18-24 age bracket, and the remaining two are between 25 and 35 years old. In terms of gender, 14 (resp. 9) of the participants self identify as female (resp. male)  and 2 self-identify as ``other". Our only criteria for selecting participants was that they have prior experience using a computer and that they do not have impaired vision. The entire user study took place over the course of three weeks.

During the user study, each participant  was asked to first complete a training session and then perform four image search tasks, two using \toolname{} and two using \baseline{}. The order of tasks, as well as which tool to use for a given task, was randomly selected.  The training session  involved completing a tutorial about both tools and performing two practice tasks, one with \toolname{} and one with \baseline{}. The users had access to the tutorial throughout the user study and were explicitly told that they could reference it whenever they wished to do so. The participants were given a total of 5 minutes to complete the practice tasks and each of the four image search tasks. Participants were told that they could end a task whenever they were satisfied with the results; however, participants opted to use all the time available to them in most cases.

In the course of the study, participants were asked to talk about their search strategies while completing each task. To aid subsequent analysis, we collected both audio and screen recordings throughout the process. Upon completion of the four tasks, the participants were asked to reflect on their experience and answer some interview questions. The total session, including the tutorial and interview, took less than 90 minutes for each participant.



\subsection{Tasks}

Given a dataset of images, the goal of each task in the user study was to identify a subset of the images matching a certain criteria. Specifically, the tasks involved the following three sets of images: 
\begin{itemize}[leftmargin=*]
    \item \emph{Transportation}: A set of 70 images of bicycles, cars, and people, mostly taken on public roads.
    \item \emph{Festival}: A set of 420 images from a music festival, comprised of images of  performances, venues, and attendees.
    \item \emph{Wedding}: A set of 352 images from a wedding, including staged photos of the wedding party and candid photos of the ceremony and reception. 
\end{itemize}
Each task targeted one of these datasets. 
Since \toolname{} is intended for use on personal images, we collected these datasets from image galleries shared on Flickr. As such, the datasets vary in size. Participants were provided with task descriptions, along with a description of the corresponding dataset. The task descriptions were as follows:
\begin{enumerate}
    \item [(0)] Find all images that contain a car and a bicycle.
    \item Find all images that contain a guitar and a microphone.
    \item Find all images that contain no people. An image contains a person if you can see any discernible part of someone’s body.
    \item Find all images where the bride is to the left of the groom. An image contains a person if their face is visible.
    \item Find all images that contain the bride but not the groom. An image contains a person if their face is visible.
\end{enumerate}
Task 0 was the practice task and involved the transportation dataset. Tasks 1 and 2 used the festival dataset, and the last two tasks involved the wedding dataset. Note that tasks 3 and 4 involve searching for \emph{particular} faces in an image. For these tasks, the participant was given example images with the bride and groom's faces.

\section{User Study Results}\label{sec:results}

\subsection{Quantitative Results}

\paragraph{Search Result Accuracy} One of the key metrics for evaluating the efficacy of each tool is \emph{accuracy of search results}. That is, within the 5 minute time limit, how close were the saved results to the ground truth? To answer this question, Table~\ref{tab:accuracy} reports the F1 score of the search results when participants use \toolname\ and \baseline. We report two different accuracy results, namely \emph{before} and \emph{after} \emph{post-processing}. To understand what we mean by this, recall that  people first interact with the underlying tool (ML model or synthesizer) to get an initial set of results, and then manually add/remove images to refine the search results before finally saving them. The columns labeled \emph{before post-processing} show the F1 score for the search result automatically generated by the tool before manual intervention.\footnote{For \toolname, this refers to the result after the user is done running the synthesizer.} As we can see, the initial search results for \toolname \ are significantly better (0.45 vs 0.76 in terms of average F1 score). Furthermore, using the Wilcoxon rank sum test, we find that these results are statically significant, with a $p$-value of less than 0.02, for all tasks. 

The columns labeled \emph{after post-processing} show the results after the users have manually refined the search results within the 5 minute time limit. Overall, the F1 scores for \toolname\ are higher compared to those of \baseline, and overall difference in F1 score is statistically significant ($p$-value of $<3.1\mathrm{e}{-7}$ for the Wilcoxon rank sum test). However, if we run the same test for each individual benchmark, we find that the result is statistically significant for only the Guitar and Microphone task and the No People task. For the Bride and Not Groom task, there was one participant who mistook a wedding guest for the bride and completed the task by searching for images containing that guest. When this outlier is removed from the dataset, the result for the Bride and Not Groom task is significant as well. A discussion of why the Bride Left of Groom task does not have a significant result is included in Section \ref{sec:limitations}.

\paragraph{Search Efficiency.} The average time per search query (i.e. the time the system takes to perform a search for a given query) for \toolname{} and \baseline{} is presented in Table~\ref{tab:accuracy}. For \baseline{}, search time is consistent across all tasks and queries. For \toolname{}, search time varies depending on what inputs the user provides. For instance, if the user provides an example image with a lot of different objects, then sketch completion will take longer, as there are more ways that the sketch could be filled in. While \toolname{}, on average, takes longer than \baseline{}, both tools are efficient enough for interactive online use.



\paragraph{Manual effort.} Another important metric for evaluating the efficacy of a tool is the amount of manual effort. That is, how many objects did the user tag, and how many images did the user have to manually add or remove before they were satisfied with the results or reached the 5 minute time limit? The use of tagging was extremely consistent. Participants used tagging only to assign names to the bride and groom in the two tasks using the wedding dataset. For instance, P14 tagged the bride and groom as ``Emily'' and ``John,'' respectively, and wrote the query ``\emph{Emily is to the left of John.}'' All but one participant who completed at least one of these tasks with \toolname{} used this strategy.

The results for other metrics of manual effort are presented in Table \ref{tab:accuracy}. In particular, for \toolname{}, we report two different numbers: (a) the total number of examples provided when using the synthesizer, and (b) the number of added/removed images to refine search results. We can take the sum of (a) and (b) to be the proxy for manual effort. The difference in manual effort between \toolname{} and \baseline{} is statistically significant for all tasks, with a $p$-value of less than 0.02. 
Note that, for all tasks, and the Bride Left of Groom task in particular, \baseline{} users manually added and removed a significant number of images in proportion to the size of the ground truth dataset. Participants using \baseline{} often resorted to extra manual efforts to add and remove images from their initial search results to achieve a higher accuracy; however, this required a greater cognitive load to complete their task: e.g., P14 mentioned \emph{``it felt like I had to basically look through every image.''}



\begin{figure}
\centering
\begin{subfigure}{.5\textwidth}
  \centering
  \includegraphics[scale=0.27, trim=50 20 50 50, clip]{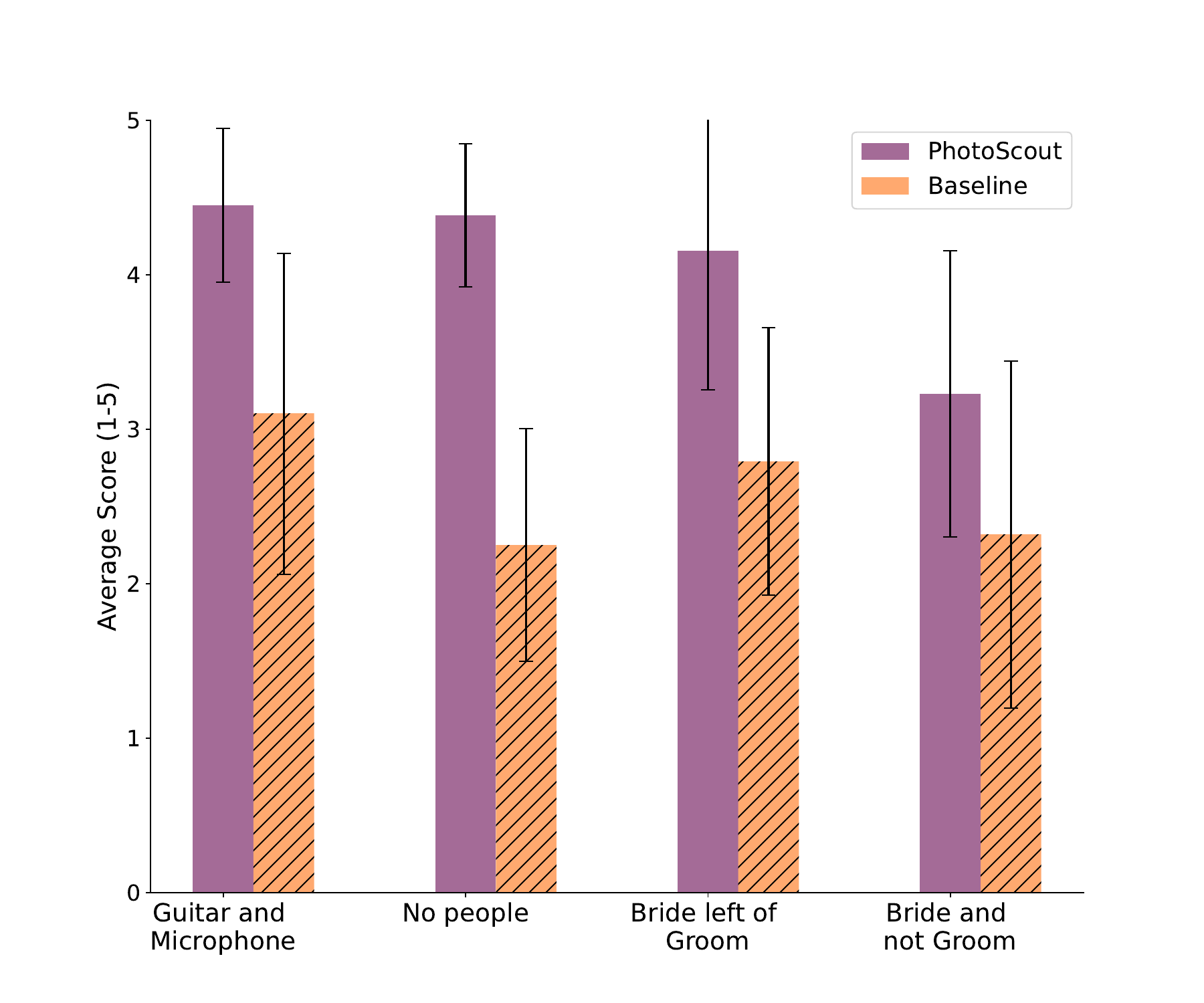}
  \label{fig:ease}
  \caption{User-reported ease of use.}
\end{subfigure}%
\begin{subfigure}{.5\textwidth}
  \centering
  \includegraphics[scale=0.27, trim=50 20 50 50, clip]{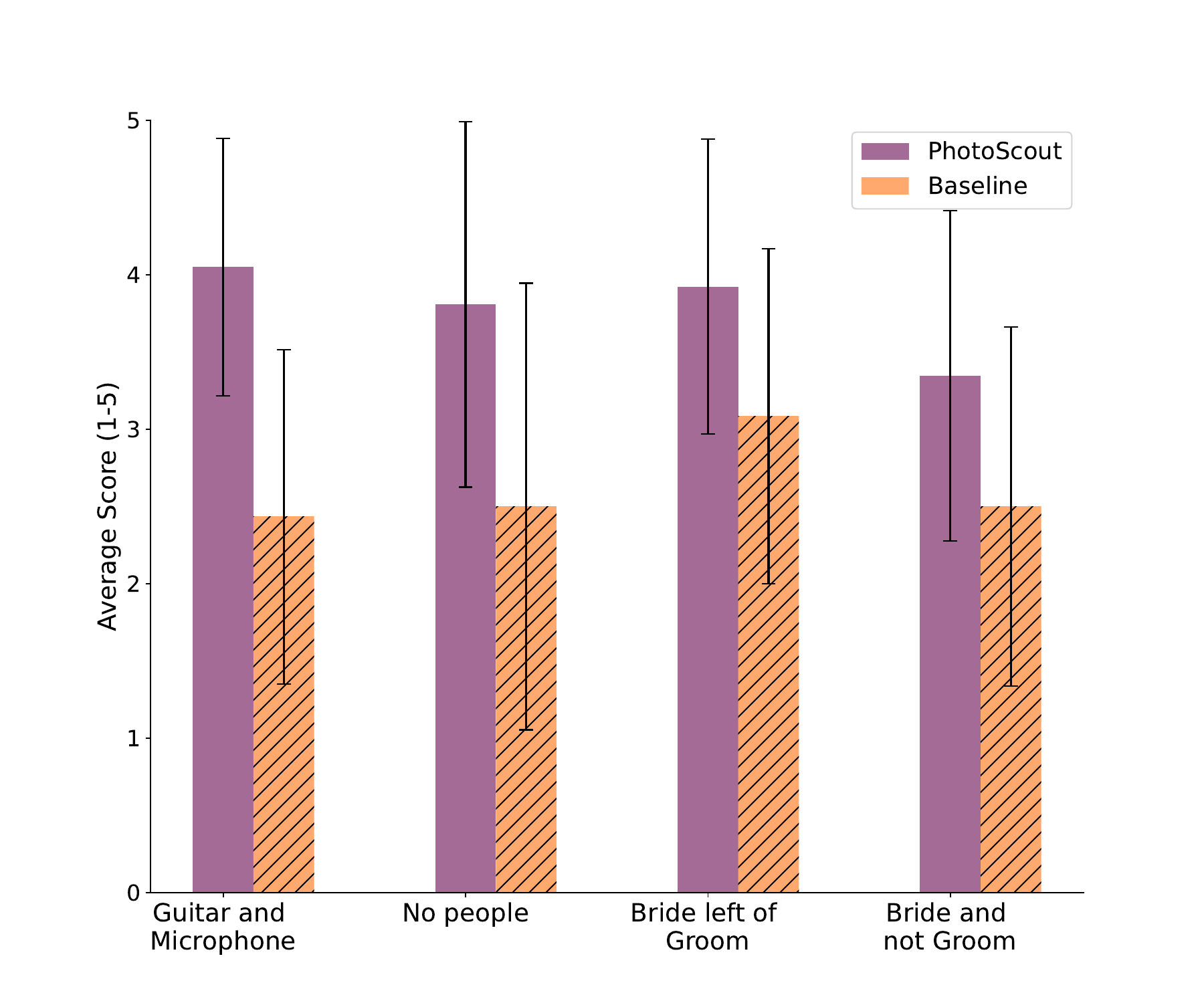}
  \label{fig:confidence}
  \caption{User-reported confidence in results.}
\end{subfigure}
\caption{Results of post-task questionnaire, with standard deviation.}
\Description{Two bar graphs listing the task on the X-axis and the average score (1-5) on the Y-axis. For each task, there is one bar for \toolname{} and one for \baseline{}. One bar graph is for question 1 and one is for question 2. In each case, the average score for \toolname{} is higher than for \baseline{}.}
\label{fig:questionnaire}
\end{figure}

\paragraph{Task Questionnaire} For the last part of our quantitative study, we analyze the results of the questionnaire that each participant was asked to complete \emph{after} finishing a task. Specifically, participants were asked the following questions upon completion of each task:
\begin{enumerate}[leftmargin=*]
    \item On a scale of 1-5, with 5 being very easy and 1 being not easy at all, how easy was it to complete the task using this tool?
    \item On a scale of 1-5, with 5 being very confident and 1 being not at all confident, how confident are you that your results are correct? ``Correct'' means that all of your saved images match the task, and none of the unsaved images match the task.
\end{enumerate}
Figure \ref{fig:questionnaire} summarizes the results of this questionnaire. Across all tasks, participants gave an average score of 4.0 for \toolname{} and 2.7 for \baseline{} on question 1, and an average score of 3.8 for \toolname{} and 2.6 for \baseline{} on question 2. For question 1, the difference in scores between \toolname{} and \baseline{} was statistically significant for all tasks, with a $p$-values less than 0.03. For question 2, the difference in scores was significant for the Guitar and Microphone task and the No People task.  

We also asked participants for qualitative input on each question.
When answering question 1 (ease of use), some participants noted that \toolname{} \emph{``has a steeper learning curve''} than \baseline{} due to its more sophisticated search features, but that \emph{``once you have done the setup, the results it gives are pretty accurate''} (P1). Further, when answering question 2 (confidence in results), some participants noted that they had confidence in their results with \baseline{} \emph{because} of the manual effort they had expended going through the images themselves. Despite these aspects working in \baseline{}'s favor, the scores for each question are consistently higher for \toolname{} than for \baseline{}. 
 
\begin{figure}
    \begin{table}[H]
    \caption{A quantitative comparison of \toolname{} (abbr. \textsc{P}) and \baseline{} (abbr. \textsc{C}). \# Ground Truth lists the number of images in the ground truth dataset of each task. \# Solved lists the number of participants who were assigned with each tool for each task. Avg. F1 Score Before and Avg. F1 Score After list, respectively, the average F1 score of the search results before and after performing post-processing (i.e. manually adding and removing images from the search results) with each task and tool. Manual Effort lists the average number of images post-processed (i.e. added and removed from search results) for each task, and, in the case of \toolname{}, the number of images selected as examples.}
\footnotesize
\begin{tabular}{|c|c|c|c|c|c|c|c|c|c|c|c|c|}
    \hline
     { \makecell{ \\ Task \\ Description}}  & \makecell{ \\ \# Ground \\ Truth} & \multicolumn{2}{c|}{ \makecell{\# Assigned}} & \multicolumn{2}{c|}
     {\makecell{ Avg. F1 Score \\  Before}}
 & \multicolumn{2}{c|}{\makecell{  Avg. F1 Score \\ After}} & \multicolumn{2}{c|}{\makecell{Avg. Time \\ Per Query (s)}} & \multicolumn{2}{c|}{\makecell{Manual Effort - \\ \toolname{}}} & \makecell{Manual Effort - \\ \baseline{}} \\
     & & \multicolumn{1}{c}{\sc {P}} & \sc {C} & \multicolumn{1}{c}{\sc {P}} & \sc {C} & \multicolumn{1}{c}{\sc {P}} & \sc {C} & \multicolumn{1}{c}{\sc {P}} & \sc {C} & \multicolumn{1}{c}{\makecell{Avg. \\ \# Examples}} & \makecell{Avg. \# Post- \\ processed} & \makecell{Avg. \# Post-\\processed} \\ 
     \hline
     \makecell{Guitar and \\ Microphone} & 63 & 10 & 14 & 0.82 & 0.58 & 0.84 & 0.63 & 1.79 & 0.09 & 3.7 & 6.1 & 28.2 \\
     \hline
     \makecell{No \\ People} & 24 & 13 & 12 & 0.78 & 0.29 & 0.77 & 0.66 & 2.08 & 0.08 & 5.1 & 4.2 & 21.4   \\
     \hline
     \makecell{Bride Left \\ of Groom} & 42 & 13 & 12 & 0.77 & 0.47 & 0.78 & 0.66 & 2.07 & 0.08 & 5.0 & 4.2 & 38.4 \\
     \hline
     \makecell{Bride and \\ Not Groom} & 40 & 11 & 13 & 0.67 & 0.43 & 0.68 & 0.54 & 2.30 & 0.08 & 4.9 & 5.8 & 21.7 \\
     \hline
     \bf { Overall } & - & \bf { 49 } &  \bf {49} & \bf {0.76} & \bf {0.45} & \bf {0.82} & \bf {0.61} & \bf{2.11} & \bf {0.08} & \bf {4.8} & \bf {5.0} & \bf {27.6} \\
     \hline
\end{tabular}
\label{tab:accuracy}
\end{table}
\end{figure}



\subsection{Qualitative Results}

We conducted a semi-structured interview about the participants' experiences using \toolname{} and \baseline{}. We asked participants about their search strategies and results using both tools. In addition, we instructed participants to think aloud while completing each task, kept notes on comments that participants made. One of the authors coded participants' responses to each question and comments on each task, and two authors reviewed and discussed the results collaboratively.  We report the following key findings:

\paragraph{KF1: \toolname's synthesis-based search procedure makes structured image search easier and more efficient} 16 of the 25 participants reported that they thought their results using \toolname{} were more accurate than their results using \baseline{}. Out of the other 9 participants, only one said that their results with \baseline{} were more accurate; the other 8 were unsure. Participants expressed confidence in their results with \toolname: \emph{“[\toolname] was actually really good at getting what I asked. I think [\toolname] was pretty trustworthy overall”} (P21). Similarly, they expressed a lack of trust in \baseline{}: \emph{``I don’t know, I just didn’t have that much faith in [\baseline]''} (P5).

In particular, participants observed that \toolname{} was better than \baseline{} at finding images that were consistent with logical and positional elements of their queries. When completing the No People task with \baseline{}, P9 noted, \emph{“I put no people in the search bar, and it gave me a bunch of images with people.”} Many participants who used \baseline{} for this task developed a strategy of finding one image without people, and searching for similar images. This strategy allowed them to find certain types of images without people (e.g. closeup images of signage at the festival), but caused them to miss other types of images that were not visually similar (e.g. photos of venues before performances had taken place). By contrast, participants who used \toolname{} could efficiently write a text query, add a few example images, and see a set of accurate search results matching the logical intention of their query. 

Similarly, when completing the Bride Left of Groom task, P18 said \emph{“[\baseline] doesn’t seem to know its lefts and rights that well.”} Participants using \baseline{} were able to find images containing the bride and groom without much difficulty, but finding images where the bride and groom were oriented correctly could only be accomplished through manually filtering. Meanwhile, participants using \toolname{} could use a text query and example images to specify that they only wanted photos were the bride is to the left of the groom, and saw results that reflected this intent. P12 stated, \emph{“I noticed that [\toolname] is more functional when it comes to relational statements”}.

Interestingly, most participants did \emph{not} make use of the natural language explanations of the search results in \toolname{}. Only 2 out of 25 participants reported that they found the explanations useful, and many participants did not notice the explanations, even though the tutorial pointed out this feature. While we cannot determine exactly why participants did not make use of NL explanations, we can conclude that this feature had little to do with participants' confidence in \toolname{}'s search results. Future work could explore alternative methods of explaining search results to the user. One such method allows users to visualize why a particular image appeared, or did not appear, in the search results. This visualization could include annotations and/or text that highlight the parts of an image that match or do not match the query. Several participants noted the potential utility of this feature when reviewing their search results.

\paragraph{KF2: Example images convey information that text alone cannot} 
22 out of 25 participants indicated that example images provided additional information that they could not convey in text. P7 said \emph{“[Examples] can describe what you’re looking for better than text.... A picture is worth a thousand words.”} P15 noted, \emph{“I like that I was able to provide example images, because it helped me clarify [my intent].”} Participants noticed that example images and text queries worked synergistically: \emph{“I gave more specific text queries in [\toolname], because I could back them up with examples”} (P25). By contrast, some participants noted that \baseline{} required more general text queries: \emph{“I tried to give [\baseline] as little ambiguity as possible”} (P4). 

During the study, example images clarified ambiguous or erroneous text queries. For instance, when completing the Guitar and Microphone task with \toolname{}, many participants made the text query ``guitar and microphone.'' Based on this text alone, it is unclear whether the user wants all images containing a guitar and a microphone, or all images containing a guitar and all images containing a microphone. A negative example image containing just one of these objects quickly resolves this ambiguity. In another instance, when completing the No People task with \toolname{}, P14 made the text query ``music festival containing no people.'' The term ``music festival'' in this query was unnecessary (as all images in the dataset were from a music festival) and could have added noise to the search results. However, because the participant had already added a set of example images for the task, \toolname{} figured out that this part of the query was extraneous, and output images containing no people. 

\paragraph{KF3: Selecting example images is an intuitive process} Every participant utilized example images when completing tasks with \toolname{}, and selecting positive and negative example images was an easy process for most tasks. Usually, the participant quickly found positive and negative example images by scanning through the full image dataset. In some instances, participants made an initial text query, and then selected example images from the smaller set of preliminary search results. 

During the tutorial, we explained what positive and negative examples were, but did not offer any insight into what makes a good or bad example image. Even so, during the study 11 out of the 25 participants noted that they intentionally selected diverse example images. For instance, P10 noted, \emph{``for negative examples, I chose things that could be confusing.''} Similarly, P7 said, \emph{``I tried to find edge cases with one image that was totally different,''} and that especially for negative examples \emph{``I tried to find sort of tricky cases where an image was almost correct.''} 

This strategy likely helps to produce correct results in \toolname{}, as the example images will filter out synthesized programs that are almost correct but are missing one key component. Even though participants were not given any information about the underlying search procedure, they independently found an effective strategy for selecting example images. This behavior suggests that example images are an intuitive addition to text-based image search. 

\subsection{Description of Failure Cases}\label{sec:limitations}

\paragraph{Limitations of object detector} 
Because \toolname{} performs image retrieval by executing neuro-symbolic queries, its performance is dependent upon the accuracy of the underlying neural models for object detection. If an image contains a particular object present in the user's query, but \toolname{} does not detect that object (e.g. because it is partially obscured), then that image will not appear in the search results.

This issue is more apparent if the object detector does not work well on the images that the user selects as positive and negative examples. In particular, because \toolname{} synthesizes a program that matches all positive examples and rejects all negative ones, \toolname{} may fail to synthesize \emph{any} programs if the object detector misclassifies relevant objects in the example images. 
Indeed, this limitation of \toolname{} proved problematic in the Bride Left of Groom task. Several participants selected an example image of the bride and groom dancing, where only the back of the bride's head is visible. While participants could easily infer that this person was the bride, \toolname{} did not classify her correctly. Hence, \toolname{} could not generate a program that matched both the user's text query and this example image, and the user was prompted to adjust their query. A common response was for participants to then add more example images in an attempt to correct this error. However, they would continue to get poor results as long as they had any example images where relevant objects were misclassified by the object detector.

As a result, participants sometimes felt more frustrated when using \toolname{} than when using \baseline{}: 8 out of 25 participants noted instances where \toolname{} should have detected an object but did not. In some cases, this caused participants to lose trust in \toolname{}. During the Bride Left of Groom task, P5 noted, \emph{``it decreases my confidence to know that [\toolname] misclassified the face I was looking for.''} When completing the Guitar and Microphone task, P8 said, \emph{``it’s annoying when [\toolname] doesn’t recognize a microphone in an image.''} 

\paragraph{Limitations of LLM} It is also the case that \toolname{}'s framework may fail to output results when GPT is unable to produce a program sketch from the user's text query. If the user provides a query that is very dissimilar from any of the example text queries in the prompt provided to GPT, then the output programs may fail to parse. This was the case when P22 made the text query \emph{``solo images of anna''} during the Bride and Not Groom task (where they had already tagged the bride as ``anna''). If this happens, the user will see no search results and will be prompted to adjust their query. 

\paragraph{Inspiration for future work} An interesting direction for future work is to explore interaction models that balance the \emph{structure} of \toolname{} with the \emph{flexibility} of \baseline{}. \baseline{} will also fail to detect objects, and often misinterprets text queries. However, \baseline{} is designed for similarity-based search queries, and does not extract any hard constraints from queries. As such, users will almost always get \emph{some} results from any query they provide to \baseline{}. Even if those results are inaccurate, users may feel more encouraged to continue trying other queries or to edit their results manually. One user suggested that \toolname{} could allow users to edit image labels in cases where the object detector is incorrect. Several other users reported that they would like a fusion of the two tools, wherein they could explore the dataset with open-ended text or image queries in a separate panel, without having to adjust the text query and example images that would determine the hard constraints of their task. 


\section{Conclusion}

We have presented \toolname{}, a new multi-modal synthesis-based interface for automating image search tasks. With \toolname{}, users provide natural language descriptions of their search tasks, then interactively select example images and tag objects to refine their search. Our approach uses an LLM to synthesize program sketches in a neuro-symbolic DSL and then grounds those sketches using a PBE approach. We have evaluated our proposed approach by conducting a user study with 25 participants, wherein users completed image search tasks with \toolname{} and a deep learning-based image search tool. We found that participants performed tasks  more accurately and with less manual work using \toolname{}.

\begin{acks}                            
  This material is based upon work supported by the
  \grantsponsor{GS100000001}{National Science
    Foundation}{http://dx.doi.org/10.13039/100000001} under Grant
  No.~\grantnum{GS100000001}{nnnnnnn} and Grant
  No.~\grantnum{GS100000001}{mmmmmmm}.  Any opinions, findings, and
  conclusions or recommendations expressed in this material are those
  of the author and do not necessarily reflect the views of the
  National Science Foundation.
\end{acks}

\bibliographystyle{ACM-Reference-Format}
\bibliography{main}

\pagebreak
\appendix
\section{Interview Questions}

The list of questions we asked participants in the user study during the semi-structured interview is as follows. Note that, during the study, we referred to \baseline{} as Tool A and \toolname{} as Tool B.

\begin{enumerate}
    \item What is your strategy to formulate text queries with image search? Do you have different strategies for Tool A versus Tool B?
    \item In Tool B, you also have to provide example images. What do you think of providing examples as a way to convey your intent? Does it provide information that you cannot convey in text? 
    \item What information do you think is easier to convey using text query versus using examples?
    \item What is your strategy for selecting example images?
    \item After you provide the initial input, both tools return a set of images. How do you check if the tool finds all images you want?
    \item If the initial results are not correct, how do you proceed? What’s your preference with the following methods: restarting your search with a new query, manually adding or removing images, or add new example images? What’s your effort required for different approaches? How do you combine these strategies?
    \item In general, do you think your submitted results are correct? Does this vary between Tool A and Tool B? \item Are you more concerned that your results have incorrect images, or that there are correct images that you are missing?
    \item If you are were not provided with these tools for today’s tasks, how would you complete the tasks if I just give you all images and task descriptions?
    \item Are there any features that you wished these tools had?
\end{enumerate}

\section{GPT Prompt}

Recall that, during synthesis, we prompt GPT with a list of (text, program) pairs, along with the user's text query, and ask it to return a program. The list of (text, program) we use in the prompt is as follows:

\begin{enumerate}
    \item Input: There is a tree in the image. \\ 
    Output: $\exists x. \textsf{HasType}(x, \textsf{Tree})$
    \item Input: The image contains a chair and a table. \\
    Output: $\exists x. \exists y. \textsf{HasType}(x, \textsf{Chair}) \wedge \textsf{HasType}(y, \textsf{Table})$
    \item Input: The image contains a chair to the left of a table. \\
    Output: $\exists x. \exists y. \textsf{HasType}(x, \textsf{Chair}) \wedge \textsf{HasType}(y, \textsf{Table}) \wedge \textsf{HasRelation}(x, y, \textsf{Left})$
    \item Input: All faces do not have eyes open. \\
    Output: $\forall x. \textsf{HasType}(x, \textsf{Face}) \rightarrow \neg \textsf{HasProperty}(x, EyesOpen)$
    \item Input: The image contains a cat inside a box. \\
    Output: $\exists x. \exists y. \textsf{HasType}(x, \textsf{Cat}) \wedge \textsf{HasType}(y, \textsf{Box}) \wedge \textsf{HasRelation}(x, y, \textsf{Inside})$  
    \item Input: Jane is in the image and everyone is smiling. \\ 
    Output: $\exists x. \textsf{HasType}(x, \textsf{Jane}) \wedge \forall y. \textsf{HasType}(y, face) \rightarrow \textsf{HasProperty}(y, \textsf{Smiling})$
    \item Input: The image contains a face that is smiling, and has their eyes open. \\ 
    Output: $\exists x. \textsf{HasProperty}(x, \textsf{Smiling}) \wedge \textsf{HasProperty}(x, \textsf{EyesOpen})$
    \item Input: Every person is next to a cat. \\
    Output: $\forall x. \textsf{HasType}(x, \textsf{Person}) \rightarrow \exists y. \textsf{HasType}(y, \textsf{Cat}) \wedge \textsf{HasRelation}(x, y, \textsf{NextTo})$
\end{enumerate}

\end{document}